\documentclass[showkeys,superscriptaddress,floatfix,prd,10pt,aps]{revtex4-2}
\usepackage{graphicx,epstopdf}
\usepackage[caption=false]{subfig}
\usepackage{appendix}
\usepackage[T1]{fontenc}
\usepackage{lmodern}
\usepackage[dvipsnames,x11names]{xcolor}
\usepackage[colorlinks=true,linkcolor=NavyBlue,citecolor=ForestGreen,urlcolor=NavyBlue]{hyperref}
\usepackage[sort&compress]{natbib}
\usepackage{lipsum}
\usepackage{morefloats}
\usepackage[pdf]{pstricks}
\usepackage{amsmath}
\usepackage{amssymb}
\usepackage{amsfonts}
\usepackage{rotating}
\usepackage{cancel}
\usepackage{mathtools}
\usepackage{bbm}
\usepackage{dsfont}
\usepackage{bbold}
\usepackage{multirow}
\usepackage{ulem}
\usepackage{physics}
\usepackage{orcidlink}
\usepackage{colortbl}
\usepackage{dcolumn}% Align table columns on decimal point
\usepackage{bm}% bold math
\usepackage{color,soul}

\def\xslash{x\!\!\!\slash }

\def\beq{\begin{equation}}
\def\eeq{\end{equation}}
\def\bea{\begin{eqnarray}}
\def\eea{\end{eqnarray}}
\def\beeq{\begin{eqnarray}}
\def\eeeq{\end{eqnarray}}

\def\vel{\left|}
\def\ver{\right|}

\def\ba{\begin{array}}
\def\ea{\end{array}}

\def\xis0{{\Xi^{*0}}}

\def\g5{\gamma_5}

\begin{document}

\title{Electromagnetic properties of vector doubly charmed tetraquark states }
\author{U. \"{O}zdem\orcidlink{0000-0002-1907-2894}}%
\email[]{ulasozdem@aydin.edu.tr}
\affiliation{ Health Services Vocational School of Higher Education, Istanbul Aydin University, Sefakoy-Kucukcekmece, 34295 Istanbul, T\"{u}rkiye}
\author{K. Azizi\orcidlink{0000-0003-3741-2167}}
\email[]{kazem.azizi@ut.ac.ir}
\affiliation{Department of Physics, University of Tehran, North Karegar Avenue, Tehran
14395-547, Iran}
\affiliation{Department of Physics, Do\v{g}u\c{s} University, Dudullu-\"{U}mraniye, 34775
Istanbul,  T\"{u}rkiye}
\affiliation{Department of Physics and Technical Sciences, Western Caspian University, Baku AZ 1001, Azerbaijan}

\date{\today}
 
\begin{abstract}
We conduct a systematic study of the electromagnetic properties of multiquark systems with undetermined internal structures. Motivated by the  recent observation of the $T_{cc}^+$ state, we apply the light-cone version of the  QCD sum rule method to extract the magnetic dipole moments of several possible doubly-charmed vector tetraquark states. When analyzing the magnetic dipole moment of these states, they are modeled to have the diquark-antidiquark configurations. The magnetic dipole moments for the members  are extracted as 
$ \mu_{T_{cc \bar{u} \bar{d}}} = 1.17^{+0.44}_{-0.32} \, \mu_N$, 
$ \mu_{T_{cc \bar{u} \bar{s}}} = 1.35^{+0.50}_{-0.37} \, \mu_N$,
$  \mu_{T_{cc \bar{d} \bar{s}}} = -2.69^{+1.02}_{-0.75} \, \mu_N$,
$ \mu_{T_{cc \bar{u} \bar{u}}}  = 1.33^{+0.56}_{-0.40} \, \mu_N$,  
$ \mu_{T_{cc \bar{d} \bar{d}}}  = 1.41^{+0.57}_{-0.43} \, \mu_N$ and 
$ \mu_{T_{cc \bar{s} \bar{s}}}  = 1.44^{+0.53}_{-0.41} \, \mu_N$. 
Comparing the results obtained for the magnetic dipole moments of the $T_{cc \bar{u} \bar{d}}$  state with the $T_{cc \bar{u} \bar{s}}$ state, the $U$-symmetry is seen to be broken at about $\%15$, while for the $T_{cc \bar{d} \bar{d}}$ and $T_{cc \bar{s} \bar{s}}$ states, this symmetry is minimally broken.  
The obtained results may be useful  to determine the true nature of these new interesting states.
\end{abstract}
%\keywords{Doubly-charmed vector tetraquarks, electromagnetic form factors,  magnetic dipole moment,  QCD light-cone sum rules}

\maketitle

\section{Introduction}\label{introduction}
Investigation of the various physical properties of multiquark systems constitutes one of the main directions of research in hadron physics. Among these systems, tetraquarks occupy a special place, since  the two main  revolutionary discoveries on multiquark systems belong to these states: the discovery of $ X(3872) $  by Belle collaboration \cite{Belle:2003nnu} in 2003  and  the discovery of $T^+_{cc}(3875)$ by LHCb Collaboration \cite{LHCb:2021vvq,LHCb:2021auc} in 2021. The first discovery was the observation of the first multiquark system  and the second one was the observation of the first doubly charmed exotic meson: the previously  discovered states were hidden heavy flavored states. At  present, we have plenty number of exotic states discovered by different experiments or have been introduced as possible candidates. Despite the good experimental and theoretical progress on the exotic states, their internal structure and quark-gluon configuration are not clear yet~\cite{Esposito:2014rxa,Esposito:2016noz,Olsen:2017bmm,Lebed:2016hpi,Nielsen:2009uh,Brambilla:2019esw,Agaev:2020zad,Chen:2016qju,Ali:2017jda,Guo:2017jvc,Liu:2019zoy,Yang:2020atz,Dong:2021juy,Dong:2021bvy,Meng:2022ozq, Chen:2022asf} and more investigations are needed. 

The discovered  $T^+_{cc}(3875)$ by the LHCb collaboration was an axial-vector tetraquark  with the quark content $cc\overline{u}\overline{d}$ and spin-parity $J^P = 1^+$  in the $D^{0}D^{0}\pi ^{+}$ invariant mass distribution as a narrow peak. Its mass lies just a little bit  below the $D^{0}D^{\ast
}(2010)^{+}$ threshold. The two-meson $D^{0}D^{\ast }{}^{+}$ threshold has a mass of  $3875.1~\mathrm{MeV}$, whereas $T_{cc}^{+}$ has the mass
$m_{\exp }=3875.1~\mathrm{MeV}+\delta m_{\exp }$ with $\delta m_{\exp }=-273\pm 61\pm 5_{-14}^{+11}~\mathrm{KeV}$. It has also  a very narrow width $\Gamma =410\pm 165\pm 43_{-38}^{+18}~\mathrm{KeV}$,  making this particle  the  longest-living exotic meson discovered so far. Because of its unique feature, the $T^+_{cc}$ and doubly-heavy tetraquark states at all  have been under intense investigations  via various models and approaches~\cite{Du:2012wp,Qin:2020zlg,Feijoo:2021ppq,Deng:2021gnb,Yan:2021wdl,Wang:2021yld,Huang:2021urd,Agaev:2021vur,Chen:2021tnn,Jin:2021cxj,Ling:2021bir,Hu:2021gdg,Chen:2021cfl,Albaladejo:2021vln,Abreu:2021jwm,Du:2021zzh,Dai:2021vgf,Wang:2021ajy,Meng:2021jnw,Fleming:2021wmk,Xin:2021wcr,Ren:2021dsi,Albuquerque:2022weq, Azizi:2021aib,Ozdem:2021hmk,Kim:2022mpa,Abreu:2022lfy,Agaev:2022ast,Deng:2021gnb,Ren:2021dsi,Albuquerque:2022weq,Deng:2021gnb,Dai:2022ulk,Dai:2021wxi,Karliner:2017qjm,Eichten:2017ffp,Cheng:2020wxa,Braaten:2020nwp, Meng:2020knc,Dias:2011mi,Navarra:2007yw,Gao:2020ogo,Agaev:2020mqq,Agaev:2020zag,Agaev:2020dba,Agaev:2019lwh,Agaev:2018khe,Aliev:2021dgx,Mohanta:2020eed,Ke:2022vsi,Ozdem:2022yhi,Agaev:2022vhq, Wang:2022jop, Qin:2022nof, Wu:2022gie, Abreu:2022sra,Dai:2023mxm,Li:2023hpk,Dai:2023cyo,Wang:2023ovj,Liu:2023ckj,Lei:2023ttd,Albuquerque:2023rrf,Azizi:2023gzv,Sonnenschein:2024rzw,Mutuk:2023oyz,Mutuk:2024vzv}.

It is very natural to search for the possible $T_{cc}$ states of different spin-parities other than $J^P = 1^+$ now both in the experiment and theory. Among them, the vector state with $J^P = 1^-$ can be very interesting  both  from the spectroscopic and electromagnetic properties points of view. It is of a great importance to exactly determine the properties of vector $T_{cc}$ and compare the results with those of its axial vector partner. From spectroscopic analyses, it is expected that the vector state lies above   the related two-meson thresholds \cite{Du:2012wp,Albuquerque:2023rrf}. This means that the vector state is unstable and  strongly decays to the two-meson states. This situation makes the vector $T_{cc}$  be very different than the axial one. The electromagnetic properties of the vector state is also expected to be different than the axial state. Theoretical investigations of different properties of the vector state and the obtained results can help experimental groups in their search for the vector $T_{cc}$ in the experiment. 
There are a few studies in the literature where the spectroscopic parameters of the vector doubly-charmed tetraquark states have been investigated in~\cite{Du:2012wp,Wang:2017dtg,Kim:2022mpa,Albuquerque:2023rrf,Sakai:2023syt}.
In \cite{Du:2012wp}, the authors developed a systematic approach to investigating potential doubly-heavy states with 
possible spin-parities by constructing possible tetraquark interpolating operators without derivatives. Their predictions indicate that there are no bound states in the doubly-charmed sectors. 
In \cite{Albuquerque:2023rrf}, the mass and coupling values of the vector states were obtained at the next-to-leading order (NLO) using the two possible interpolating currents constructed in \cite{Du:2012wp}. It is shown that the mass values obtained from this study are approximately 200-300 MeV lower than the mass values in \cite{Du:2012wp}. 
In Ref.~\cite{Wang:2017dtg}, the author constructed axialvector-diquark–axialvector-antidiquark type currents to interpolate vector and other spin-parities doubly-charmed tetraquark states. These currents were then used to study these states with QCD sum rules, with the operator product expansion carried out up to vacuum condensates of dimension 10 in a consistent manner. 
In \cite{Kim:2022mpa}, the mass of the vector and other possible double-heavy states are studied in the chiral-diquark picture. Based on these findings, it was concluded that there are no bound doubly-charmed tetraquark states. 
In \cite{Sakai:2023syt}, the masses of the vector and other possible doubly-heavy states were calculated within the framework of the one-boson exchange model, considering their molecular structure. The results demonstrated that only doubly-charmed axial states can be stable. 
It should be noted, however, that their existence and stability are contingent upon the specific model in question, at least for the time being. Further theoretical investigation would be useful in clarifying this issue.

The masses and residues of doubly-charmed tetraquark states are available~\cite{Du:2012wp,Albuquerque:2023rrf}, allowing us to use them as inputs to discover the electromagnetic parameters of the vector states of different configurations. The physical quantities related to the electromagnetic properties of hadrons are useful parameters to reveal the nature and inner structures of specially the exotic states. The electromagnetic form factors (FFs)  and the   resultant multipole moments help us to investigate the charge and magnetism distributions inside the hadrons. This also allows us to know where and how the quark-antiquark (the valence and sea) and gluons are distributed inside the volume of the hadrons and to gain useful information about the geometric shapes and electric and magnetic radii of the hadrons. In the case of the proton, now, we see that there is a puzzle regarding the electromagnetic, mass,  and mechanical radii of this relatively well-known particle. Hence, investigation of the interaction of exotic states with non-zero spin and charge with the photon can provide useful information on many aspects of the exotic hadrons. Calculation of such kinds of non-perturbative objects requires some reliable non-perturbative approaches. Among them is QCD sum rule formalism  \cite{Shifman:1978bx,Shifman:1978by}, which is one of the powerful and predictive non-perturbative approaches in hadron physics.   
Note that the magnetic dipole moment of the axial-vector  $T_{cc}$ states was investigated in \cite{Azizi:2023gzv} using the light-cone  QCD sum rules  \cite{Chernyak:1990ag, Braun:1988qv, Balitsky:1989ry}. However, as we mentioned above, there are important differences between the nature, internal structure and  decay properties of the vector and axial-vector states that are resulted from different currents interpolating these states. The axial state was discovered in the experiment, so we hope theoretical calculations of different related parameters  will help and motivate  experimental investigations of the vector $T_{cc}$ state.

The rest of this article is arranged as follows. In Sect. \ref{formalism}, we apply the QCD light-cone sum rules to calculate the MDMs of doubly-charmed vector tetraquark states.  Section \ref{numerical} is dedicated to the numeric analyses of the attained  sum rules in the previous section. The final section contains   a brief discussion and concluding remarks.

% \begin{widetext}

 \section{Formalism} \label{formalism}

 \begin{widetext}

As is well known, the QCD light-cone sum rule is a widely used method to probe the hadronic properties and has proven to be a robust non-perturbative technique to extract different physical quantities in the non-perturbative regime of QCD. For deriving the magnetic dipole moments, we take into account an appropriate  correlator in the  presence of a weak external electromagnetic  field:

\begin{equation}
 \label{edmn01}
\Pi _{\mu \nu }(p,q)=i\int d^{4}xe^{ip\cdot x}\langle 0|\mathcal{T}\{J_{\mu}(x)
J_{\nu }^{ \dagger }(0)\}|0\rangle_{\gamma}, 
\end{equation}%
%where 
where the sub-index $\gamma$ indicates the weak background electromagnetic field and $J_{\mu}(x)$ represents  the  current interpolating  the vector  $T_{cc}$ states with the total angular momentum-parity $ J^{P} = 1^{-}$. To determine the magnetic dipole moments of doubly-charmed vector tetraquark states, we apply diquark-antidiquark interpolating currents as shown below

\begin{eqnarray}\label{curr1}
&  J^1_{\mu}(x)= \big[c^{aT}(x)C  c_b(x)][\bar q_1^a(x) \gamma_\mu C \bar q_2^b(x)],
\\
 %\nonumber\\
   &  J^2_{\mu}(x)= \big[c^{aT}(x)C \gamma_\mu \gamma_5 c_b(x)][\bar q^a(x) \gamma_5 C \bar q^b(x)],
     \end{eqnarray}
where $q$ denotes the $u$, $d$ and $s$-quarks; the $q_1$ is the $u$ or $d$-quark,  the $q_2$ is the $d$ or $s$-quark. While $q_1$ is the $u$-quark, $q_2$ can be the  $d$ or  $s$-quark. If $q_1$ is the $d$-quark, then $q_2$ can be  only the $s$-quark. 
In principle, various interpolating currents of the molecular and compact tetraquark forms couple to  these states~\cite{Du:2012wp}. Calculations up to next-to-leading order (NLO) on the spectroscopic parameters show that the above currents lead to more reliable results \cite{Albuquerque:2023rrf}. Hence, using these  currents,  the values of the mass and current coupling as the main input parameters in the calculations of the electromagnetic properties of the states under study are available with higher accuracy, allowing us to use them in extraction of the magnetic dipole moments of the vector $T_{cc}$ states. These parameters are not available for other possible interpolating currents.

%%%%%%%%%%%%%%%%%%%%%%%%%%%%%%%%%%%%

The calculation of magnetic dipole moments starts with the analysis of the correlation function with the help of the  hadronic parameters. To achieve this, we inject a full set of hadronic states, with the same quantum numbers as the  currents under study, into  the correlation function. By following this procedure and performing the resultant four-integrals, we obtain
 \begin{align}
\label{edmn04}
\Pi_{\mu\nu}^{Had} (p,q) &= {\frac{\langle 0 \mid J_\mu (x) \mid
T_{cc}(p, \varepsilon^i) \rangle}{p^2 - m_{T_{cc}}^2}} %\nonumber\\
%&
\langle T_{cc}(p, \varepsilon^i) \mid T_{cc}(p+q, \varepsilon^f) \rangle_\gamma 
%\nonumber\\
%&
\frac{\langle T_{cc}(p+q,\varepsilon^f) \mid {J^\dagger}_\nu (0) \mid 0 \rangle}{(p+q)^2 - m_{T_{cc}}^2} \nonumber\\
&+\mbox{higher states}\,.
\end{align}

To continue the calculations, we parameterize  the matrix elements appearing in  the above equation in terms of different quantities:
\begin{align}
%\label{edmn05}
\langle 0 \mid J_\mu(x) \mid T_{cc}(p,\varepsilon^i) \rangle &= \lambda_{T_{cc}} \varepsilon_\mu^i\,,\\
%%%%%%%%%%%%%%%%%%%%%%%%%%%
\langle T_{cc}(p,\varepsilon^i) \mid  T_{cc} (p+q,\varepsilon^{f})\rangle_\gamma &= - \varepsilon^\gamma (\varepsilon^{i})^\alpha (\varepsilon^{f})^\beta \bigg\{ G_1(Q^2) (2p+q)_\gamma ~g_{\alpha\beta} 
%\nonumber\\
%& \times  
%\nonumber\\
%& \times 
+ G_2(Q^2) ( g_{\gamma\beta}~ q_\alpha -  g_{\gamma\alpha}~ q_\beta) 
\nonumber\\ &- \frac{1}{2 m_{T_{cc}}^2} G_3(Q^2)~ (2p+q)_\gamma 
%\nonumber\\
%& \times 
q_\alpha q_\beta  \bigg\},\label{edmn06}
\end{align}
where $\varepsilon^\gamma$ is the polarization vector  of the photon, $\varepsilon^i$ and $\varepsilon^f$ indicate the polarization vectors of the
initial and final doubly-charmed vector  states, and  $\lambda_{T_{cc}}$ is residue or current coupling of the doubly-charmed tetraquark state $ T_{cc} $.  Here,  $G_i(Q^2)$ with $i=1,2$, and $3$ are the electromagnetic form factors at $ Q^2=-q^2 $.

The physical or  phenomenological  side of the correlator is acquired  by making use of  the Eqs. (\ref{edmn04})-(\ref{edmn06}) as follows:

\begin{align}
\label{edmn09}
 \Pi_{\mu\nu}^{Had}(p,q) &=  \frac{\varepsilon_\rho \, \lambda_{T_{cc}}^2}{ [m_{T_{cc}}^2 - (p+q)^2][m_{T_{cc}}^2 - p^2]}
 \Big\{ G_2 (Q^2) %\nonumber\\
% & \times 
 \Big(q_\mu g_{\rho\nu} - q_\nu g_{\rho\mu} -
\frac{p_\nu}{m_{T_{cc}}^2}  \big(q_\mu p_\rho - \frac{1}{2}
Q^2 g_{\mu\rho}\big) 
  \nonumber\\
 &  
 +
\frac{(p+q)_\mu}{m_{T_{cc}}^2}  \big(q_\nu (p+q)_\rho+ \frac{1}{2}
Q^2 g_{\nu\rho}\big) %\nonumber\\
%&
-  
\frac{(p+q)_\mu p_\nu p_\rho}{m_{T_{cc}}^4} \, Q^2
\Big)
%\nonumber\\
%&
+ \cdots \Big\},
\end{align}
where $ \cdots $ here stands for the contributions of the higher states and continuum as well as other Lorentz structures that are not used to extract the required FFs. As is seen, we kept only the form factor $ G_2(Q^2) $ and the  corresponding Lorentz structure. We need only this form factor which is called  magnetic FF, 
\begin{align}
\label{edmn07}
&F_M(Q^2) = G_2(Q^2)\,,
\end{align}
 the static limit of which is used to extract the  MDM, $\mu_{T_{cc}}$:
\begin{align}
\label{edmn08}
&\mu_{T_{cc}} = \frac{ e}{2\, m_{T_{cc}}} \,F_M(Q^2=0).
\end{align}

The physical side led us to define the MDM of the state under study. Now, we need to evaluate the correlator in high energies and short distances in terms of QCD parameters called the QCD representation. To this end,  we insert the explicit forms of the  currents in terms of the quark fields  into the correlator and contract the corresponding fields of the heavy and light quarks by means of  Wick's theorem. When this is done, we obtain the following  representations,
\begin{eqnarray}
\Pi _{\mu \nu }^{\mathrm{QCD}}(p,q)&=&i\int d^{4}xe^{ip\cdot x} \, \langle 0 \mid  
%\nonumber\\
%&& 
\Big\{
\mathrm{Tr}\Big[  S_{c}^{bb^{\prime }}(x)   \widetilde S_{c}^{aa^{\prime }}(-x)\Big]
\mathrm{Tr}\Big[ \gamma _{\mu }\widetilde S_{q_2}^{b^{\prime }b}(-x) \gamma _{\nu } S_{q_1}^{a^{\prime }a}(-x)\Big]    
\nonumber\\
&&
-\mathrm{Tr}\Big[ S_{c}^{ba^{\prime }}(x)  \widetilde S_{c}^{ab^{\prime }}(-x)\Big]
\mathrm{Tr}\Big[ \gamma _{\mu } \widetilde S_{q_2}^{b^{\prime }b}(-x) \gamma _{\nu }S_{q_1}^{a^{\prime }a}(-x)\Big]    
\nonumber\\
&&-
\mathrm{Tr}\Big[  S_{c}^{bb^{\prime }}(x)   \widetilde S_{c}^{aa^{\prime }}(-x)\Big]
\mathrm{Tr}\Big[ \gamma _{\mu } \widetilde S_{q_1q_2}^{a^{\prime }b}(-x) \gamma _{\nu }S_{q_2 q_1}^{b^{\prime }a}(-x)\Big] 
\nonumber\\
&&
+\mathrm{Tr}\Big[  S_{c}^{ba^{\prime }}(x)    \widetilde S_{c}^{ab^{\prime }}(-x)\Big]
\mathrm{Tr}\Big[ \gamma _{\mu } \widetilde S_{q_1q_2}^{a^{\prime }b}(-x) \gamma _{\nu } S_{q_2q_1}^{b^{\prime }a}(-x)\Big] 
\Big\} \mid 0 \rangle_{\gamma} ,  \label{QCDSide1}
\end{eqnarray}%
for the $J_\mu^1$  and 

\begin{eqnarray}
\Pi _{\mu \nu }^{\mathrm{QCD}}(p,q)&=&i\int d^{4}xe^{ip\cdot x} \, \langle 0 \mid  
%\nonumber\\
%&& 
\Big\{
\mathrm{Tr}\Big[ \gamma _{\mu } \gamma _{5}S_{c}^{bb^{\prime
}}(x) \gamma _{5} \gamma _{\nu } \widetilde S_{c}^{aa^{\prime }}(-x)\Big]
\mathrm{Tr}\Big[ \gamma _{5} \widetilde S_{q}^{b^{\prime }b}(-x)\gamma _{5}S_{q}^{a^{\prime }a}(-x)\Big]    
\nonumber\\
&&
-\mathrm{Tr}\Big[ \gamma _{\mu } \gamma _{5} S_{c}^{ba^{\prime
}}(x)  \gamma _{5} \gamma _{\nu } \widetilde S_{c}^{ab^{\prime }}(-x)\Big]
\mathrm{Tr}\Big[ \gamma _{5} \widetilde S_{q}^{b^{\prime }b}(-x)\gamma _{5}S_{q}^{a^{\prime }a}(-x)\Big]    
\nonumber\\
&&-
\mathrm{Tr}\Big[ \gamma _{\mu } \gamma _{5}S_{c}^{bb^{\prime
}}(x) \gamma _{5} \gamma _{\nu } \widetilde S_{c}^{aa^{\prime }}(-x)\Big]
\mathrm{Tr}\Big[ \gamma _{5} \widetilde S_{q}^{a^{\prime }b}(-x)\gamma _{5}S_{q}^{b^{\prime }a}(-x)\Big] 
\nonumber\\
&&
+\mathrm{Tr}\Big[ \gamma _{\mu } \gamma _{5} S_{c}^{ba^{\prime
}}(x)  \gamma _{5} \gamma _{\nu } \widetilde S_{c}^{ab^{\prime }}(-x)\Big]
\mathrm{Tr}\Big[ \gamma _{5} \widetilde S_{q}^{a^{\prime }b}(-x)\gamma _{5}S_{q}^{b^{\prime }a}(-x)\Big] 
\Big\} \mid 0 \rangle_{\gamma} ,  \label{QCDSide2}
\end{eqnarray}%
for the $J_\mu^2$ current considered in the present study. Here, $S_{q}(x)$ and $S_{c}(x)$ denote the light and charm-quark propagators, which are written as
%During our calculations, we utilize the x-space expressions for the light and heavy-quark propagators,
\begin{align}
\label{edmn12}
S_{q}(x)&=i \frac{{\xslash}}{2\pi ^{2}x^{4}} 
- \frac{\langle \bar qq \rangle }{12} \Big(1-i\frac{m_{q} \xslash}{4}   \Big)
- \frac{ \langle \bar qq \rangle }{192}m_0^2 x^2   %\nonumber\\
%& \times 
\Big(1-i\frac{m_{q} \xslash}{6}   \Big)
-\frac {i g_s }{32 \pi^2 x^2} ~G^{\mu \nu} (x) \Big[\rlap/{x} 
\sigma_{\mu \nu} +  \sigma_{\mu \nu} \rlap/{x}
 \Big]+\cdots,
\end{align}
%and
%%
\begin{align}
\label{edmn13}
S_{c}(x)&=\frac{m_{c}^{2}}{4 \pi^{2}} \Bigg[ \frac{K_{1}\Big(m_{c}\sqrt{-x^{2}}\Big) }{\sqrt{-x^{2}}}
+i\frac{{\xslash}~K_{2}\Big( m_{c}\sqrt{-x^{2}}\Big)}
{(\sqrt{-x^{2}})^{2}}\Bigg] %\nonumber\\
%&
-\frac{g_{s}m_{c}}{16\pi ^{2}} \int_0^1 dv\, G^{\mu \nu }(vx)\Bigg[ \big(\sigma _{\mu \nu }{\xslash}
  +{\xslash}\sigma _{\mu \nu }\big)\nonumber\\
  &\times \frac{K_{1}\Big( m_{c}\sqrt{-x^{2}}\Big) }{\sqrt{-x^{2}}}
+2\sigma_{\mu \nu }K_{0}\Big( m_{c}\sqrt{-x^{2}}\Big)\Bigg]+\cdots,
\end{align}%
where $\langle \bar qq \rangle$ is the light-quark  condensate, $m_q$ is the light-quark mass,   $ m_0^2= \langle 0 \mid \bar  q\, g_s\, \sigma_{\mu\nu}\, G^{\mu\nu}\, q \mid 0 \rangle / \langle \bar qq \rangle $,  
$G^{\mu\nu}$ is the gluon field strength tensor, $\sigma_{\mu\nu}= \frac{i}{2}[\gamma_\mu,\gamma_\nu]$, $m_c$ is the charm-quark mass, the $K_i$ are modified Bessel functions of the second kind and $v$ is the line variable. 

The  QCD side of the correlator contains both the  perturbative and non-perturbative pieces giving contributions to the problem under study.  
The perturbative contributions are related to the perturbative soft interaction of the  photon  with the light and heavy quark lines. In order to calculate such contributions,  one of the heavy/light quark propagators, having interaction with the photon,  is replaced by
\begin{align}
\label{free}
S^{pert}(x) \rightarrow \int d^4z\, S^{pert} (x-z)\,\rlap/{\!A}(z)\, S^{pert} (z)\,,
\end{align}
where $S^{pert}(x)$  stands for  the first term of the light or heavy quark propagator, and the rest  propagators that have no interaction with the photon are taken as their perturbative parts only.  
The non-perturbative part should contain non-perturbative large distance interactions of the photon with the quark lines. Such contributions are written in terms of some matrix elements like $\langle \gamma(q)\vel \bar{q}(x) \Gamma_i q(0) \ver 0\rangle$ and $\langle \gamma(q)\vel \bar{q}(x) \Gamma_i G_{\mu\nu}q(0) \ver 0\rangle$ with $\Gamma_i = \textbf{1}, \gamma_5, \gamma_\mu, i\gamma_5 \gamma_\mu, \sigma_{\mu\nu}/\sqrt{2}$, being the Dirac set.  These matrix elements can be decomposed in terms of distribution amplitudes (DAs) of the photon having different twists. The relevant matrix elements and DAs of  the photon together with the corresponding wavefunctions and entered constants all are represented/calculated in Ref. \cite{Ball:2002ps}. To achieve the non-perturbative contributions that represent the interaction of the photon with light quark lines  at a large scale, we make use of the  following replacement for one of the light quark propagators interacting with the photon
\begin{align}
\label{edmn14}
S_{\mu\nu}^{ij}(x) \rightarrow -\frac{1}{4} \big[\bar{q}^i(x) \Gamma_i q^j(0)\big]\big(\Gamma_i\big)_{\mu\nu},
\end{align}
 where  the other propagators  are considered  as their full  expressions.  These procedures produce the matrix elements discussed above. 
Adding both the resultant perturbative and non-perturbative contributions gives the QCD side of the correlation function in terms of QCD fundamental parameters as well as the  quark-gluon degrees of freedom.

Finally, we choose the gauge invariant  $(q_\mu \varepsilon_\nu- q_\nu \varepsilon_\mu)$ structure from both the hadronic and QCD sides  of the correlation function  and match its coefficients from both representations.  The contributions from the higher states and the continuum are suppressed by the Borel transformation and the continuum subtraction. As a result of the procedures we have described above, we  obtain the following  sum rules
\begin{align}\label{sonj1}
 \mu^1_{T_{cc}} &= \Delta_1 (\rm{M^2},\rm{s_0}),
 \end{align}
 \begin{align}\label{sonj2}
  \mu^2_{T_{cc}} &= \Delta_2 (\rm{M^2},\rm{s_0}),
\end{align}
where the $\mu^1_{T_{cc}}$ and $\mu^2_{T_{cc}}$ denote the magnetic dipole moments obtained for the $T_{cc \bar q_1 \bar q_2}$ and $T_{cc \bar q \bar q}$ states, respectively. The results for the $\Delta_1 (\rm{M^2},\rm{s_0})$ and $\Delta_2 (\rm{M^2},\rm{s_0})$ functions are listed in  appendix , explicitly.

\end{widetext}

\section{Numerical analysis  of the sum rules}\label{numerical}

In this section, we numerically analyze the MDMs of the vector doubly-charmed tetraquark states  obtained via the QCD light-cone sum rules   in the previous section. 
The values of the input parameters required for the numerical calculations are listed as $m_u=m_d=0$, $m_s =93.4^{+8.6}_{-3.4}\,\mbox{MeV}$, $m_c = 1.27 \pm 0.02\,$GeV, 
$f_{3\gamma}=-0.0039$~GeV$^2$~\cite{Ball:2002ps},  
$\langle \bar ss\rangle $= $0.8 \langle \bar uu\rangle$ with 
$\langle \bar uu\rangle $=$(-0.24\pm 0.01)^3\,$GeV$^3$~\cite{Ioffe:2005ym},  
$m_0^{2} = 0.8 \pm 0.1$~GeV$^2$~\cite{Ioffe:2005ym} and  $\langle g_s^2G^2\rangle = 0.88~ $GeV$^4$~\cite{Matheus:2006xi}. %and  $\chi=-2.85 \pm 0.5$~GeV$^{-2}$~\cite{Rohrwild:2007yt}. 
The numerical values of the hadronic parameters such as the masses and residues of the doubly-charmed vector tetraquark states are borrowed from Refs.~\cite{Du:2012wp,Albuquerque:2023rrf}. The DAs of the photon are among the main input parameters for the calculations of the magnetic dipole moments, we use them from the Ref. \cite{Ball:2002ps}.

In establishing the QCD light-cone sum rules, two helping parameters, $\rm{M^2}$ and $\rm{s_0}$, have been entered as previously mentioned. These parameters should be fixed according to the standard prescriptions that follow two criteria: Maximum possible pole contribution (PC)  and convergence of the operator product expansion (COPE). To fulfill these criteria, it is convenient to impose the  following  conditions, typical for the exotic states,
\begin{align}
 \mbox{PC} =\frac{\Delta_i (\rm{M^2},\rm{s_0})}{\Delta_i (\rm{M^2},\infty)} \geq  30\%,
 \\
 \nonumber\\
 \mbox{COPE} (\rm{M^2}, \rm{s_0}) =\frac{\Delta_i^{\rm{Dim 7}} (\rm{M^2},\rm{s_0})}{\Delta_i (\rm{M^2},\rm{s_0})} \leq 5\%,
 \end{align}
 where $\Delta_i^{\rm{Dim 7}} (\rm{M^2},\rm{s_0})$ denotes the last term in the OPE  of $\Delta_i (\rm{M^2},\rm{s_0})$. By these requirements, we fix the working widows of the auxiliary parameters  $\rm{M^2}$ and $\rm{s_0}$. We present these intervals for the states under study in Table \ref{parameter}. In this table, we also present the values of $ \mbox{PC} $ and $  \mbox{COPE} $ obtained from the analyses for each state.

In  Figure 1, we depict the dependencies of the MDMs of the doubly-charmed vector  tetraquark states on $\rm{M^2}$ at different fixed values of the continuum threshold $\rm{s_0}$. From this figure, it follows that the MDMs exhibit mild dependence on these helping parameters in their working intervals.

\begin{widetext}

\begin{table}[htp]
	\addtolength{\tabcolsep}{10pt}
	\caption{Working intervals of  the  $\rm{s_0}$, $\rm{M^2}$, the PC  and COPE  for the MDMs of the doubly-charmed vector tetraquark states.}
	\label{parameter}
		\begin{center}
		%\scalebox{1.0}{
\begin{tabular}{l|ccccc}
                \hline\hline
                \\
Current&~~State~~  & $\rm{s_0}$ (GeV$^2$)& 
$\rm{M^2}$ (GeV$^2$) & ~~  PC ($\%$) ~~ & ~~  COPE 
$\leq$ ($\%$) \\
\\
                                        \hline\hline
                                        \\
 &~~$cc \bar{u} \bar{d}$~~      &  $[20.5, 22.5]$ & $[4.2, 6.2]$ & $[58.3, 31.2]$ &  $2.52$  
                        \\
                        \\
 ~~$J_\mu^1$~~&~~$cc \bar{u} \bar{s}$~~     &  $[21.0, 23.0]$ & $[4.3, 6.3]$ & $[58.0, 32.3]$ &  $2.58$  \\
                        \\
&~~$cc \bar{d} \bar{s}$~~      & $[21.0, 23.0]$ & $[4.3, 6.3]$ & $[58.1, 32.4]$ & $ 2.61$   \\
\\
                                      \hline\hline
                                      \\
 &~~$cc \bar{u} \bar{u}$~~   & $[20.5, 22.5]$ & $[4.2, 6.2]$ & $[58.4, 31.4]$ & $2.71$  \\
                        \\
  ~~$J_\mu^2$~~&~~$cc \bar{d} \bar{d}$~~  & $[20.5, 22.5]$ & $[4.2, 6.2]$ & $[58.1, 32.3]$ &  $2.73$  \\
 \\
 &~~$cc \bar{s} \bar{s}$~~ & $[21.5, 23.5]$  & $[4.5, 6.5]$ & $[58.4, 32.4]$ &  $2.72$  \\
 \\
                                    \hline\hline
 \end{tabular}
%}
\end{center}
\end{table}

\end{widetext}

Having determined all the input parameters, we are ready to calculate the corresponding MDMs of the vector doubly-charmed tetraquark states. Our final results are given as
\begin{align}\label{MMres}
 \mu_{T_{cc \bar{u} \bar{d}}} &= 1.17^{+0.44}_{-0.32} \, \mu_N,~~~~~~
 \mu_{T_{cc \bar{u} \bar{u}}}  = 1.33^{+0.56}_{-0.40} \, \mu_N,
 \nonumber\\
  \mu_{T_{cc \bar{u} \bar{s}}} &= 1.35^{+0.50}_{-0.37} \, \mu_N,~~~~~~
  \mu_{T_{cc \bar{d} \bar{d}}}  = 1.41^{+0.57}_{-0.43} \, \mu_N, 
 \nonumber\\
  \mu_{T_{cc \bar{d} \bar{s}}} &= -2.69^{+1.02}_{-0.75} \, \mu_N,~~~~
  \mu_{T_{cc \bar{s} \bar{s}}}  = 1.44^{+0.53}_{-0.41} \, \mu_N,
\end{align}
 where, the variations with respect to the errors of input parameters and  the working intervals of the helping quantities $\rm{M^2}$ and $\rm{s_0}$  together with  the errors in the numerical values of the photon DAs, are the main sources of the uncertainties in the presented results.
The orders of the MDMs indicate that they are accessible in the experiment. Our analyses  also  depict that the magnetic dipole moments of $T_{cc \bar q_1 \bar q_2}$ states are governed by the light-diquarks, while those of $T_{cc \bar q \bar q}$ states are governed by the heavy-diquarks. 
Comparing the results of MDMs for different flavors leads to a conclusion on the order of  breaking of the  $U$-symmetry among the states.   Although the $U$-symmetry breaking effects have been taken into account through a non-zero $s$-quark mass and  different $s$-quark condensate, we observe that the $U$-symmetry violation in the magnetic dipole moments is negligible between  the $T_{cc \bar d \bar d}$ and $T_{cc \bar s \bar s}$  states, and is approximately $\%15$ between the $T_{cc \bar u \bar d}$ and  $T_{cc \bar u \bar s}$ states.

We should mention that the MDMs of the doubly-bottom vector tetraquarks have also been investigated together with the charmed channels, however, the results are not included in the text as  reliable sum rules could not be constructed.
In the analysis of the doubly-bottom vector tetraquarks, the fundamental requirements  of the method regarding the PC and COPE  have not been sufficiently satisfied, and reliable results could not be obtained. It should be noted that, as we previously mentioned,  the mass values for the vector doubly-bottom tetraquark were obtained using the chosen interpolating currents in~\cite{Du:2012wp,Albuquerque:2023rrf}. However, given that the physical quantities studied in the present work have different dynamics, this situation can be the case. 

As a by-product, we also calculate the quadrupole moments of the vector doubly-charmed tetraquark states. The results obtained are:
\begin{align}\label{QMres}
 \mathcal{D}_{T_{cc \bar{u} \bar{d}}} &= (-2.91^{+0.56}_{-0.79}) \times 10^{-3} \, \mbox{fm}^2,
 \nonumber\\
  \mathcal{D}_{T_{cc \bar{u} \bar{s}}} &= (-3.17^{+0.62}_{-0.55}) \times 10^{-3} \, \mbox{fm}^2, 
 \nonumber\\
  \mathcal{D}_{T_{cc \bar{d} \bar{s}}} &= (3.19^{+0.62}_{-0.56}) \times 10^{-3}\, \mbox{fm}^2.
\end{align}
 
The quadrupole moments of the $T_{cc \bar u \bar u}$, $T_{cc \bar d \bar d}$, and $T_{cc \bar s \bar s}$ states have also been calculated. Their values are very close to zero.

Before ending this section, we would like to make some comments on the importance and indirect ways of the measurements of the electromagnetic parameters of the $ T_{cc} $ states. As we previously mentioned, the  $T^+_{cc}(3875)$ with  $J^P = 1^+$ discovered by the LHCb collaboration lies just  below the $D^{0}D^{\ast}(2010)^{+}$ threshold.  This situation, prevents the fast strong dissociation of this particle  and make  it  the  longest-living exotic meson discovered so far. The long lifetime of this particle would allow  experimental groups to measure the electromagnetic properties of this particle like its magnetic dipole moment to shed light on its yet exactly undetermined nature and inner structure. The theoretical results obtained for its electromagnetic properties together with those of its possible single and double strange partners   available from the theory \cite{Azizi:2023gzv} can help the experimental groups in this respect. It is expected that  the $ T_{cc} $ with $J^P = 1^+$ is the lightest state in the  $cc\overline{u}\overline{d}$ channel. Hence, the resonances with other quantum numbers like $J^P = 1^-$ are expected to lie above   the related two-meson thresholds \cite{Du:2012wp,Albuquerque:2023rrf} making their lifetime shorter than the axial-vector state considering their possible strong decays. Although, this situation can make the measurements of its parameters difficult compared to the $ T_{cc} $ with $J^P = 1^+$, it is possible to identify the vector $ T_{cc} $ state and indirectly  measure its electromagnetic properties  by considering the recent progresses in the experiments. Any measurement on the properties of vector state will help us gain useful information on the nature and quark-gluon organization of the interesting doubly charmed states.  There are  alternative ways to indirectly determine the electromagnetic properties of the vector mesons. The first one is based on the soft photon emission of hadrons proposed in Ref.~\cite{Zakharov:1968fb}, where a procedure for the determination of the electromagnetic multipole moments is given. 
The basic idea of this procedure is that the photon carries information about  the magnetic dipole and other higher multipole moments of the particle from which it is emitted. The matrix element for the radiative process can be expressed by the energy of the photon, $E_\gamma$, as
\begin{align}
 M \sim \frac{A}{E_\gamma} + B(E_\gamma )^0 + \cdots,
\end{align}
where  $\frac{1}{E_\gamma}$, $(E_\gamma)^0$, and $\cdots$  represent the contributions coming from the electric charge, the magnetic dipole moment, and the higher multipole moments, respectively.  Therefore, the magnetic dipole moment of the  particles  under consideration can be identified through measuring the cross section or decay width of the radiative process and ignoring the small contributions of the linear/higher order terms in $E_\gamma$. The  $ \Delta^+(1232)$ resonance also has a very short lifetime, whereas the magnetic dipole moment of this resonance was obtained by utilizing the experimental data obtained in the $\gamma N \rightarrow  \Delta \rightarrow  \Delta \gamma \rightarrow \pi N \gamma $ process with the help of this technique~\cite{Pascalutsa:2004je, Pascalutsa:2005vq, Pascalutsa:2007wb}.  
 The second one is the possibility of measuring the electromagnetic multipole moments of the vector mesons in the radiative production and decays of such mesons, and it has been argued that the energy and the combined angular distributions of the radiated photons is an effective method to measure the electromagnetic multipole moments of the vector mesons~\cite{LopezCastro:1997dg}. This approach has also been used for the determination of the magnetic dipole moment of the $\rho$ meson with preliminary data from the BaBar Collaboration for the $e^+ e^- \rightarrow \pi^+ \pi^- 2 \pi^0$ process~\cite{GarciaGudino:2013alv}.  Therefore, although it is currently not possible to experimentally measure the electromagnetic properties of resonances directly, they can be obtained using the  experimental data of corresponding radiative processes indirectly.

 \begin{figure}[htp]
\centering
\subfloat[]{\includegraphics[width=0.45\textwidth]{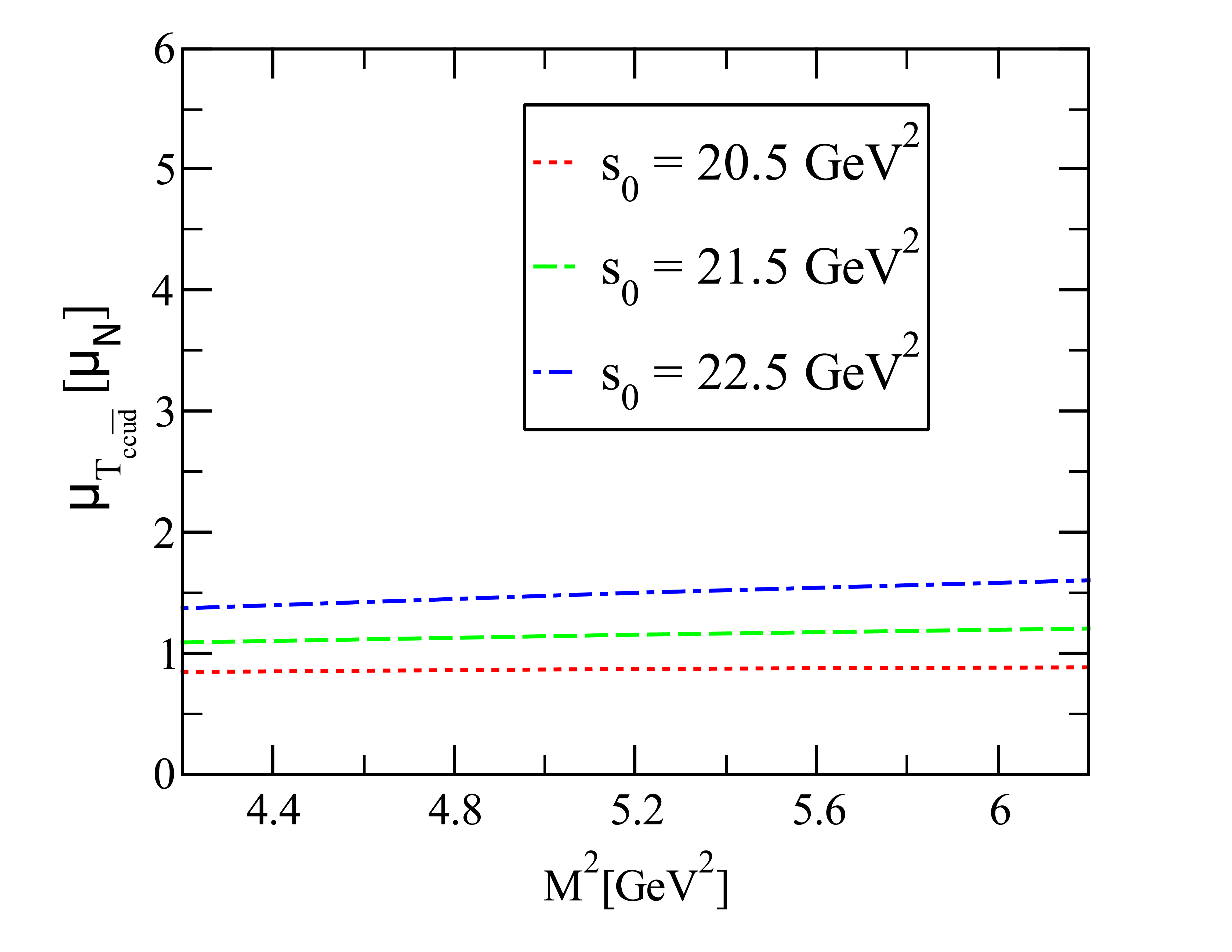}}
\subfloat[]{\includegraphics[width=0.45\textwidth]{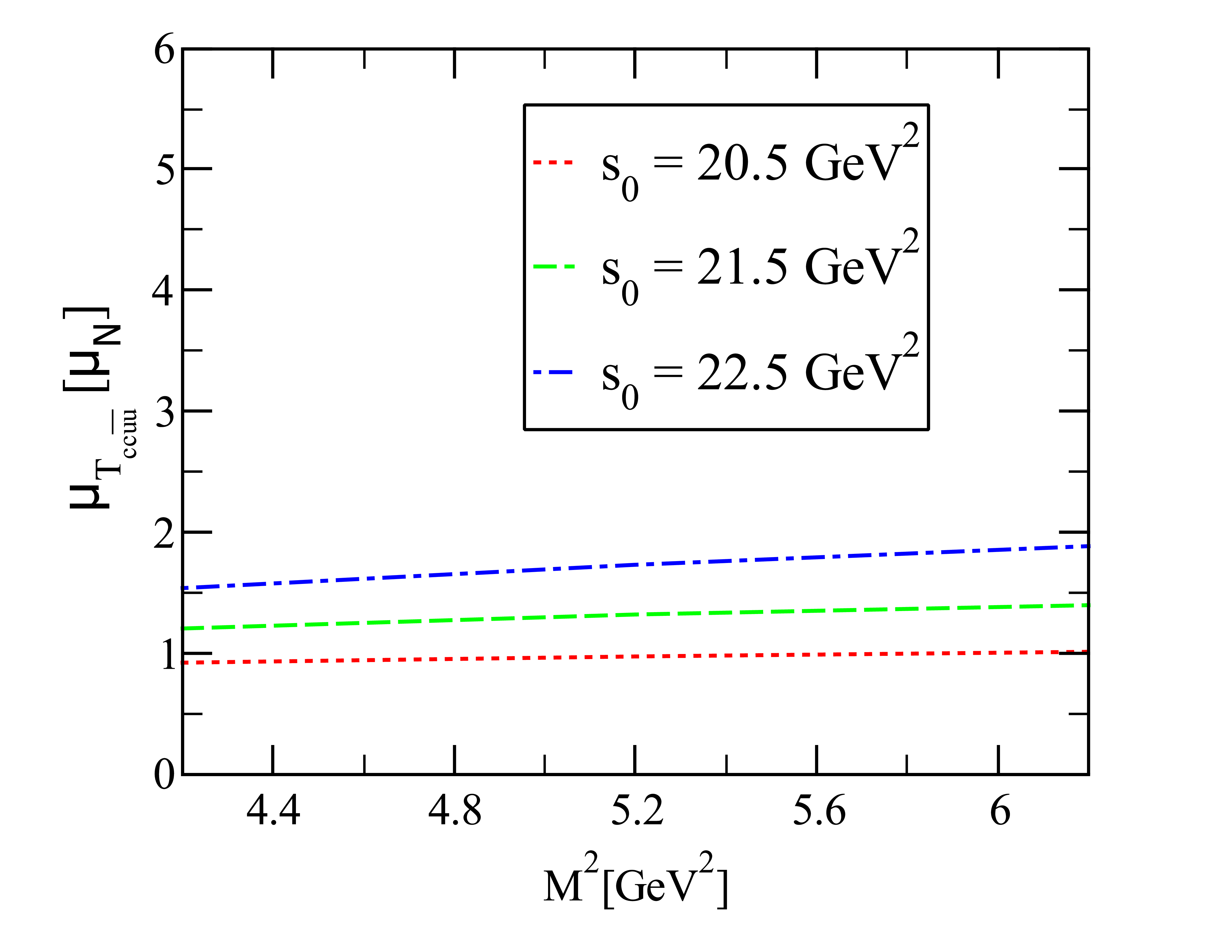}}\\
\subfloat[]{\includegraphics[width=0.45\textwidth]{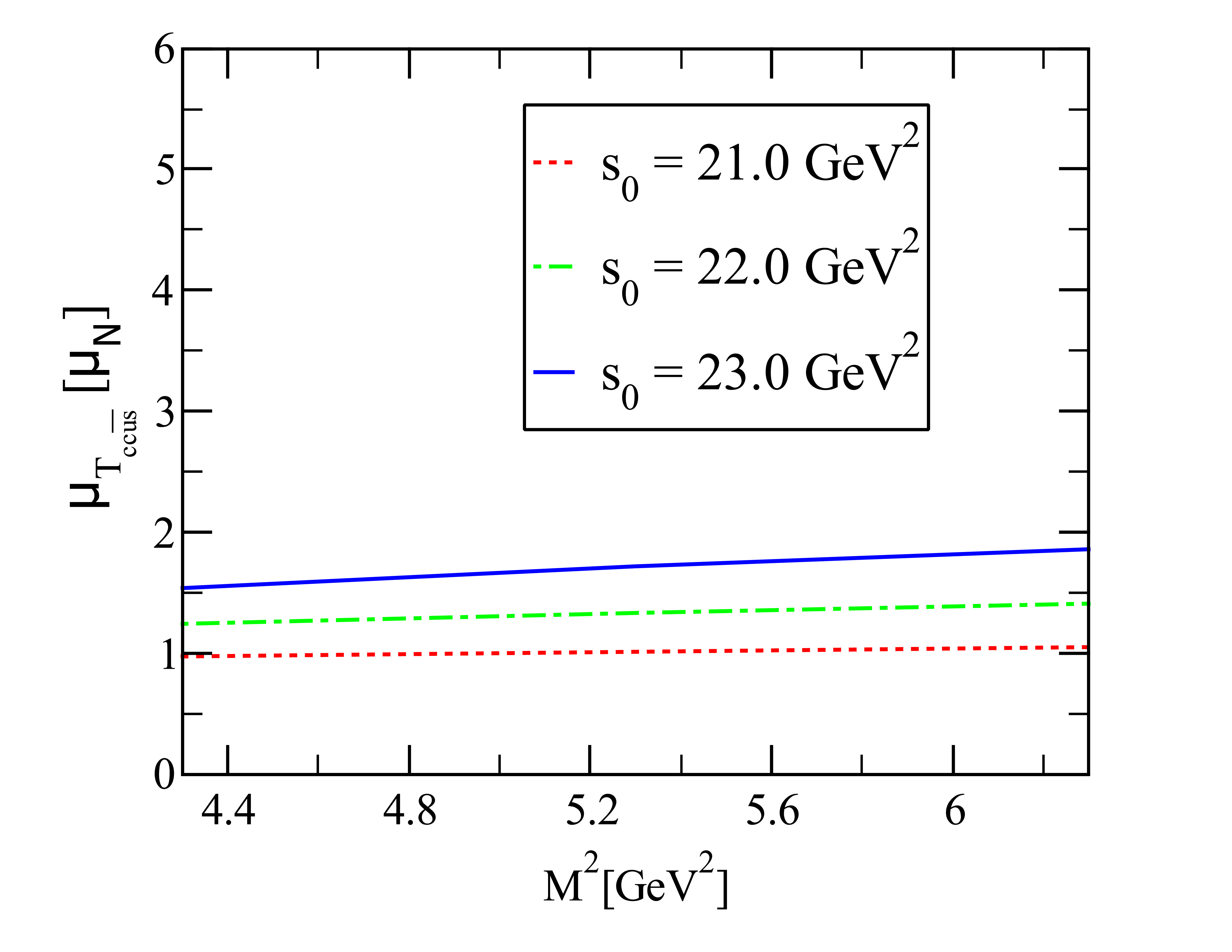}}
\subfloat[]{\includegraphics[width=0.45\textwidth]{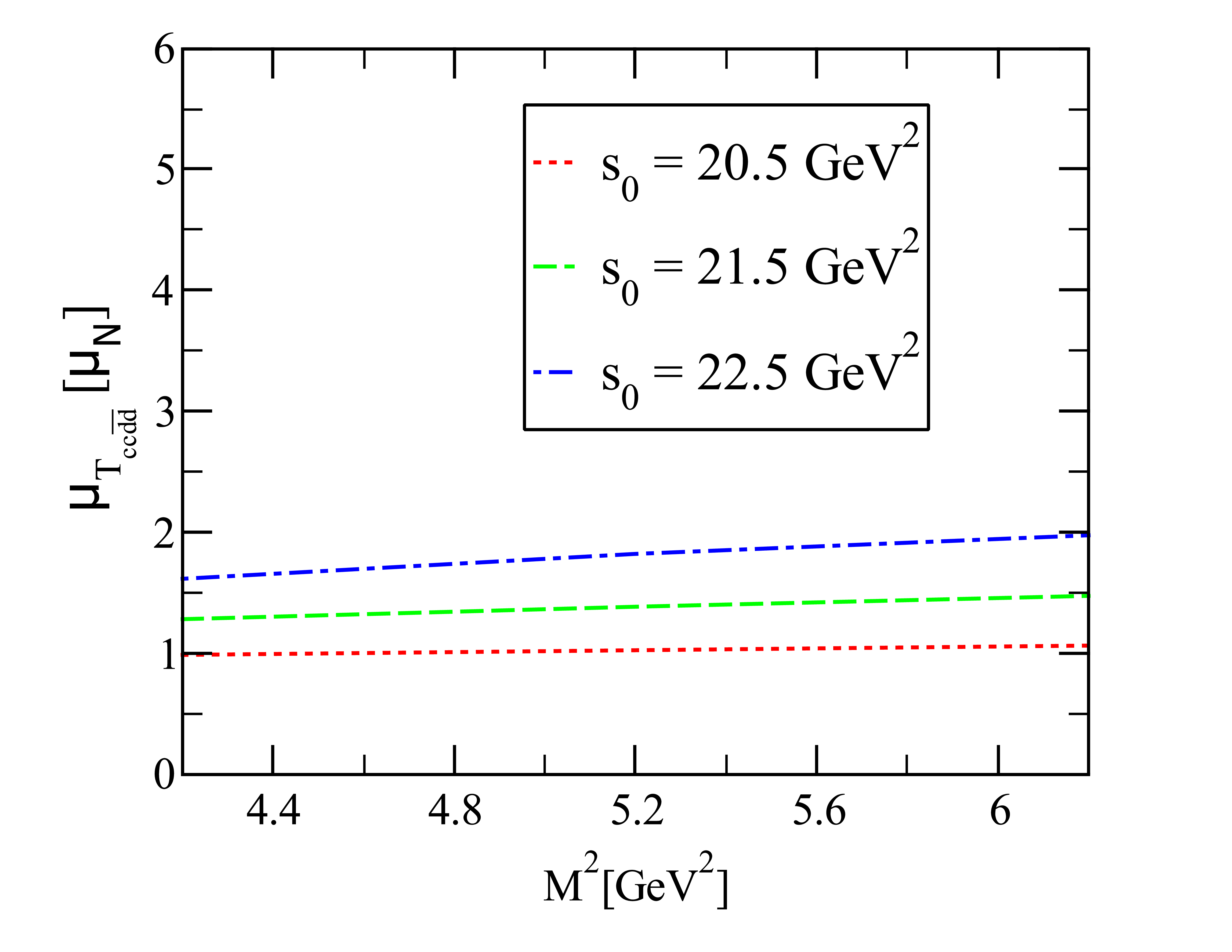}}\\
\subfloat[]{\includegraphics[width=0.45\textwidth]{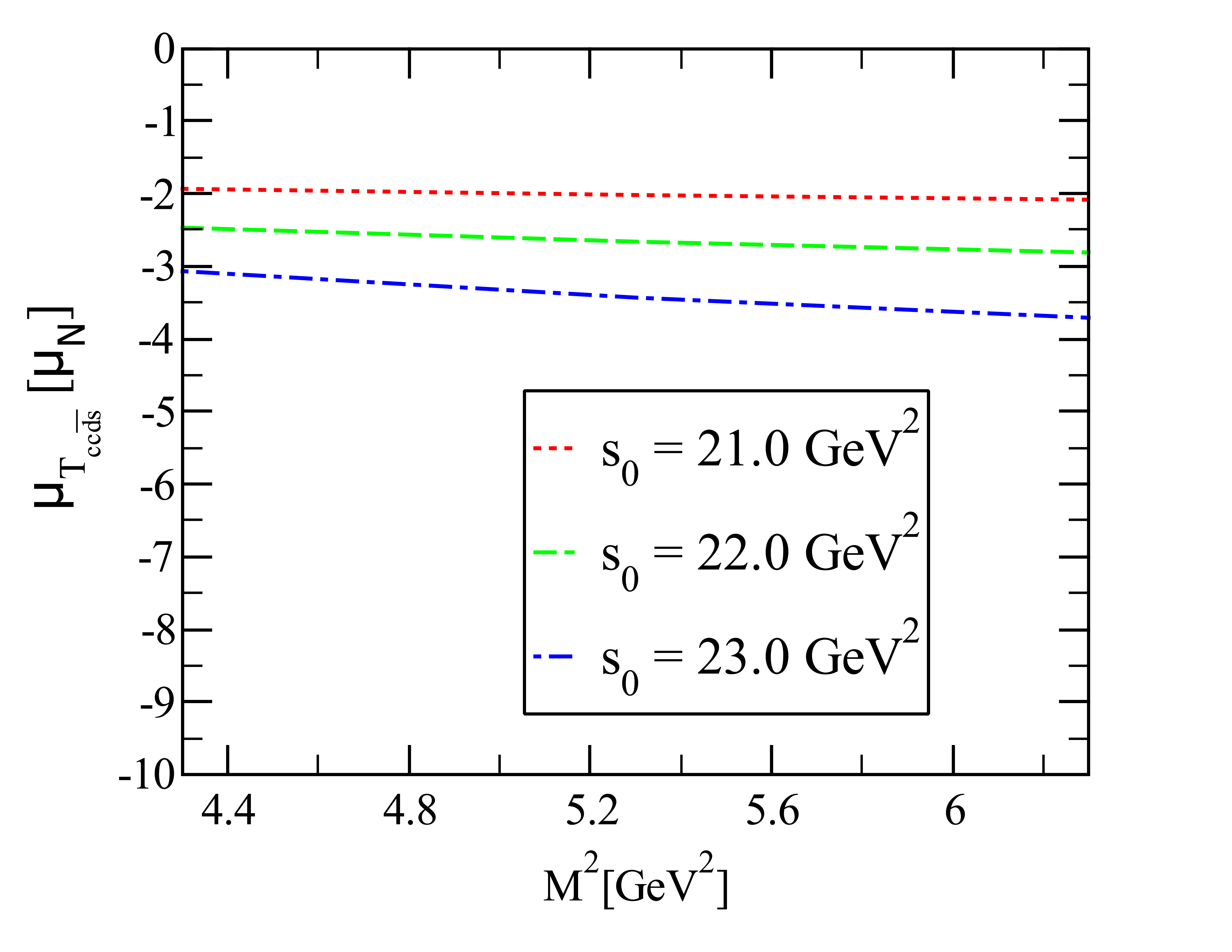}}
\subfloat[]{\includegraphics[width=0.45\textwidth]{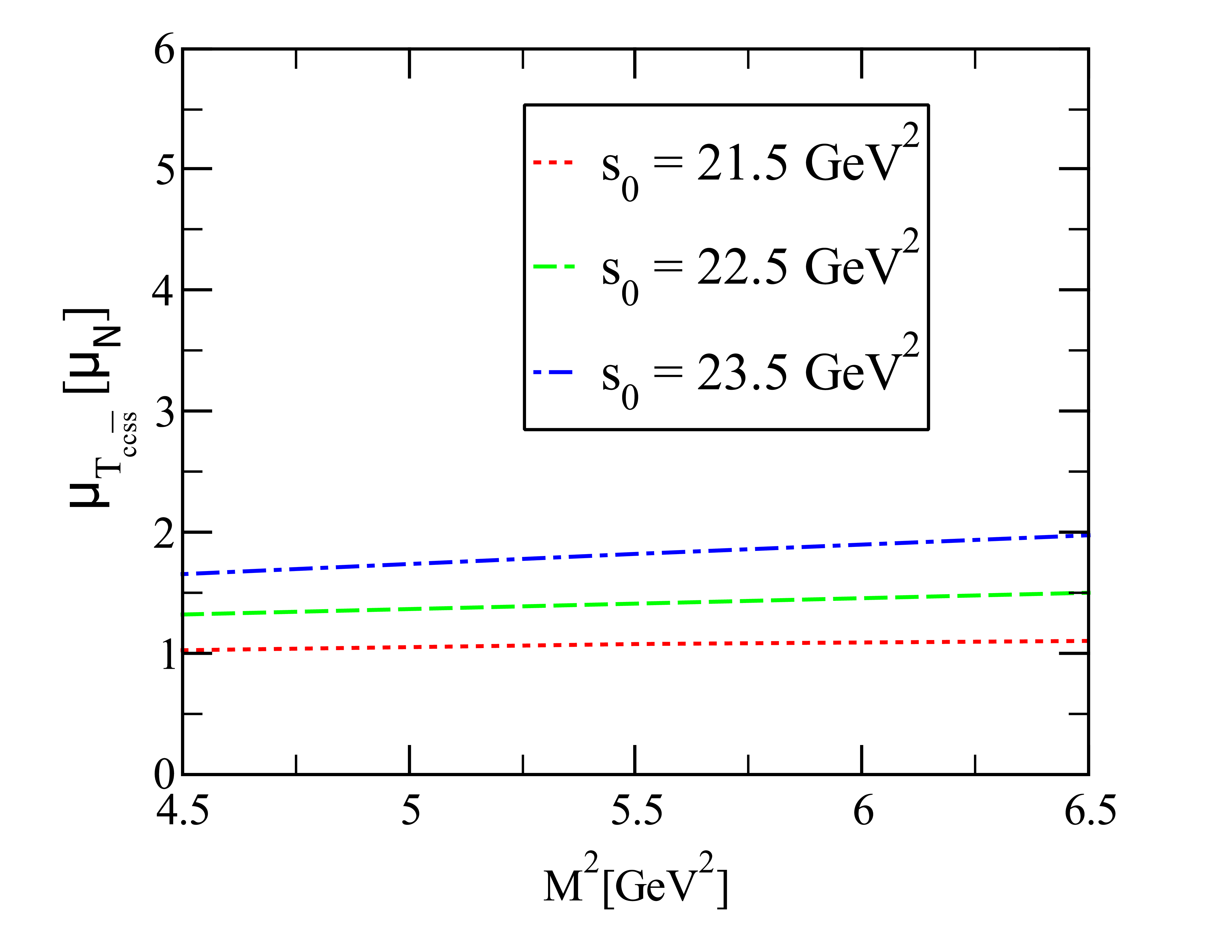}}
 \caption{Dependencies of the MDMs of the doubly-charmed vector tetraquark states on $\rm{M^2}$ at three different values of $\rm{s_0}$.}
 \label{Msqfig1}
  \end{figure}

\section{Summary and Concluding Remarks}

The experimental observation of the axial-vector  $T^+_{cc}$ state,  and the measured values of the corresponding  mass and width that give this particle   a unique place and make it a long-lived exotic state discovered ever,  has opened a new area to  study  the double-charmed tetraquark states: Scientists in this field have conducted extensive research on the nature of these states after this discovery. The studying of the electromagnetic properties of these states can help us better understand the nature, internal structure and quark-gluon configurations of these states. 

Motivated by this, we  determined the MDMs of the  possible vector doubly-charmed  states with spin-parity $ J^{P} = 1^{-}$. We  assigned a diquark-antidiquark structure to these states and considered the possible light quark contents. We used the QCD light-cone sum rule formalism to extract the values of the MDMs. The values of the presented MDMs and quadrupole moments together with the spectroscopic/decay parameters of these states increase our understanding of the nature and inner structures of these states. They may help experimental groups in the course of their search for doubly-heavy tetraquarks of different spin-parities.

 \section*{Acknowledgments}
K. Azizi is thankful to Iran National Science Foundation (INSF) for the  partial financial support provided under the elites Grant No.  4025036.

\appendix*
  
 \section{ Explicit expressions for $\Delta_1 (\rm{M^2},\rm{s_0})$ and $\Delta_2 (\rm{M^2},\rm{s_0})$ functions}
 We present  the explicit expressions of the functions $\Delta_1 (\rm{M^2},\rm{s_0})$ and $\Delta_2 (\rm{M^2},\rm{s_0})$ for the MDMs of vector doubly-charmed  tetraquark states entering the obtained sum rules:
\begin{align}
  \Delta_1 (\rm{M^2},\rm{s_0}) &= \frac{e^{\frac{m_{T_{cc}}^2}{\rm{M^2}}}}{\lambda^2_{T_{cc}}}\, \Bigg\{ \frac {(e_ {q_ 1} + 
     e_ {q_ 2}) m_c^2 } {327680 \pi^5}\Bigg[
   2 I[0, 5, 2, 1] - 5 I[0, 5, 2, 2] + 4 I[0, 5, 2, 3] - 
    I[0, 5, 2, 4] - 4 I[0, 5, 3, 1]  \nonumber\\
 &+ 6 I[0, 5, 3, 2] - 
    2 I[0, 5, 3, 3] + 2 I[0, 5, 4, 1] - I[0, 5, 4, 2] + 
    5 I[1, 4, 2, 2]- 10 I[1, 4, 2, 3]  + 5 I[1, 4, 2, 4] \nonumber\\
 & - 
    10 I[1, 4, 3, 2] + 10 I[1, 4, 3, 3] + 5 I[1, 4, 4, 2]\Bigg]\nonumber\\
  %%%%%%%%%%%%%%%%%%%%%%%%%%%%%%%%%%%%%%%%%%%%%%%%%%%%%%%%%%%%%%%%%%%%%%%%%%%%%%%%%%%%%%%%%%
 &-
 \frac{3(e_ {q_ 1} + 
     e_ {q_ 2})}{1310720 \pi^5} \Bigg[ I[0, 6, 3, 0] - 4 I[0, 6, 3, 1] + 6 I[0, 6, 3, 2] - 4 I[0, 6, 3, 3] + 
 I[0, 6, 3, 4] - 3 I[0, 6, 4, 0] \nonumber\\
 &+ 9 I[0, 6, 4, 1] - 
 9 I[0, 6, 4, 2] + 3 I[0, 6, 4, 3] + 3 I[0, 6, 5, 0] - 
 6 I[0, 6, 5, 1] + 3 I[0, 6, 5, 2] - I[0, 6, 6, 0]\nonumber\\
 & + I[0, 6, 6, 1] + 
 6 I[1, 5, 3, 1] - 18 I[1, 5, 3, 2] + 18 I[1, 5, 3, 3] - 
 6 I[1, 5, 3, 4] -18 I[1, 5, 4, 1] + 36 I[1, 5, 4, 2] \nonumber\\
 &-  
 18 I[1, 5, 4, 3] + 18 I[1, 5, 5, 1] - 18 I[1, 5, 5, 2] - 
 6 I[1, 5, 6, 1] \Bigg]\nonumber\\
  %%%%%%%%%%%%%%%%%%%%%%%%%%%%%%%%%%%%%%%%%%%%%%%%%%%%%%%%%%%%%%%%%%%%%%%%%%%%%%%%%%%%%%%%%%
 &
     -\frac {e_ {q_ 1} m_ {q_ 2} m_c^2 \langle g_s^2 G^2 \rangle \langle \bar q_1 q_1 \rangle} {36864 \pi^3}  (4 I[0, 1, 2, 0] - 3 I[0, 1, 4, 0])I_ 6[
  h_ {\gamma}]
  \nonumber\\
  %%%%%%%%%%%%%%%%%%%%%%%%%%%%%%%%%%%%%%%%%%%%%%%%%%%%%%%%%%%%%%%%%%%%%%%%%%%%%%%%%%%%%%%%%%
 &-\frac {e_ {q_ 2} m_ {q_ 1} m_c^2 \langle g_s^2 G^2 \rangle \langle \bar q_2 q_2 \rangle} {36864 \pi^3}  (4 I[0, 1, 2, 0] - 3 I[0, 1, 4, 0])I_ 6[
  h_ {\gamma}]\nonumber\\
  %%%%%%%%%%%%%%%%%%%%%%%%%%%%%%%%%%%%%%%%%%%%%%%%%%%%%%%%%%%%%%%%%%%%%%%%%%
  &-\frac {35 (e_ {q_ 1} + 
     e_ {q_ 2})  m_c^2 \langle g_s^2 G^2 \rangle f_ {3 \gamma}} {3538944 \pi^3} I[0, 2, 2, 
  0] I_ 2[\mathcal V]\nonumber\\
  %%%%%%%%%%%%%%%%%%%%%%%%%%%%%%%%%%%%%%%%%%%%%%%%%%%%%%%%%%%%%%%%%%%%%
 & +\frac {(e_ {q_ 1} + e_ {q_ 2}) m_c^2 f_ {3 \gamma}} {393216 \pi^3} I[
  0, 4, 4, 0] I_ 2[\mathcal V]
    \Bigg\},%\\
  %  \nonumber\\
\end{align}
\begin{align}
 \Delta_2 (\rm{M^2},\rm{s_0}) &= \frac{e^{\frac{m_{T_{cc}}^2}{\rm{M^2}}}}{\lambda^2_{T_{cc}}}\, \Bigg\{ 
 \frac{e_c m_c^2 }{163840 \pi^5}\Bigg[I[0, 5, 1, 2] - 2 I[0, 5, 1, 3] + I[0, 5, 1, 4] - 2 I[0, 5, 2, 2] + 
  2 I[0, 5, 2, 3] + I[0, 5, 3, 2]\Bigg]\nonumber\\
  %%%%%%%%%%%%%%%%%%%%%%%%%%%%%%%%%%%%%%%%%%%%%%%%%%%%%%%%%%%%%%%%%%%%%%%%%%%%%%%%%%%%%%%
  &+ \frac{e_c}{327680 \pi^5}\Bigg[ 2 I[0, 6, 2, 1] - 7 I[0, 6, 2, 2] + 9 I[0, 6, 2, 3] - 
 5 I[0, 6, 2, 4] + I[0, 6, 2, 5] - 6 I[0, 6, 3, 1] \nonumber\\
 &+ 
 15 I[0, 6, 3, 2] - 12 I[0, 6, 3, 3] + 3 I[0, 6, 3, 4] + 
 6 I[0, 6, 4, 1] - 9 I[0, 6, 4, 2] + 3 I[0, 6, 4, 3] - 
 2 I[0, 6, 5, 1] \nonumber\\
 &+ I[0, 6, 5, 2] + 6 I[1, 5, 2, 2] - 
 18 I[1, 5, 2, 3] + 18 I[1, 5, 2, 4] - 6 I[1, 5, 2, 5] - 
 18 I[1, 5, 3, 2] + 36 I[1, 5, 3, 3] \nonumber\\
 &- 18 I[1, 5, 3, 4] + 
 18 I[1, 5, 4, 2] - 18 I[1, 5, 4, 3] - 6 I[1, 5, 5, 2]\Bigg]\nonumber\\
  %%%%%%%%%%%%%%%%%%%%%%%%%%%%%%%%%%%%%%%%%%%%%%%%%%%%%%%%%%%%%%%%%%%%%%%%%%%%%%%%%%%%%%%%%%
 &
    +\frac {e_ {q} m_ {q} m_c^2 \langle g_s^2 G^2 \rangle \langle \bar q q \rangle} {147456 \pi^3}  I[0, 1, 2, 0] I_ 6[
  h_ {\gamma}]
  \nonumber\\
 %%%%%%%%%%%%%%%%%%%%%%%%%%%%%%%%%%%%%%%%%%%%%%%%%%%%%%%%%%%%%%
 &+\frac {e_q  m_c^2 \langle g_s^2 G^2 \rangle } {3538944 \pi^3} I[0, 3, 2, 
  0] I_ 2[\mathcal V]\nonumber\\
 %%%%%%%%%%%%%%%%%%%%%%%%%%%%%%%%%%%%%%%%%%%%%%%%%
 &+\frac {e_q m_q m_c^2 \langle \bar q q \rangle  }{3072 \pi^3} I[0, 3, 3, 0]I_4[\mathcal S]\nonumber\\
 &+\frac {e_q m_q \langle \bar q q \rangle} {32768 \pi^3} I[0, 4, 5, 0] I_2[\mathcal S]\nonumber\\
 &-\frac {e_q m_c^2  f_ {3 \gamma}} {24576 \pi^3}I[0, 4, 3, 
  0] I_ 2[\mathcal V]\nonumber\\
  &+\frac {e_q  f_ {3 \gamma}} {327680 \pi^3} I[0, 5, 5, 
  0] I_ 2[\mathcal V]
 \Bigg\},
\end{align}
where  the functions $I[n,m,l,k]$ and $I_i[\mathcal{F}]$  are
defined as:
\begin{align}
 I[n,m,l,k]&= \int_{4 m_c^2}^{\rm{s_0}} ds \int_{0}^1 dt \int_{0}^1 dw~ e^{-s/\rm{M^2}}~
 s^n\,(s-4\,m_c^2)^m\,t^l\,w^k,\nonumber\\
 %  \end{align}
 %\begin{align}
 I_1[\mathcal{F}]&=\int D_{\alpha_i} \int_0^1 dv~ \mathcal{F}(\alpha_{\bar q},\alpha_q,\alpha_g)
 \delta'(\alpha_ q +\bar v \alpha_g-u_0),\nonumber\\
  I_2[\mathcal{F}]&=\int D_{\alpha_i} \int_0^1 dv~ \mathcal{F}(\alpha_{\bar q},\alpha_q,\alpha_g)
 \delta'(\alpha_{\bar q}+ v \alpha_g-u_0),\nonumber\\
    I_3[\mathcal{F}]&=\int D_{\alpha_i} \int_0^1 dv~ \mathcal{F}(\alpha_{\bar q},\alpha_q,\alpha_g)
 \delta(\alpha_ q +\bar v \alpha_g-u_0),\nonumber\\
   I_4[\mathcal{F}]&=\int D_{\alpha_i} \int_0^1 dv~ \mathcal{F}(\alpha_{\bar q},\alpha_q,\alpha_g)
 \delta(\alpha_{\bar q}+ v \alpha_g-u_0),\nonumber\\
 %\end{align}
 %\begin{align}
   I_5[\mathcal{F}]&=\int_0^1 du~ \mathcal{F}(u)\delta'(u-u_0),\nonumber\\
 I_6[\mathcal{F}]&=\int_0^1 du~ \mathcal{F}(u),
 \end{align}
with  $\mathcal{F}$ being the corresponding photon DAs.

 \bibliography{Vector_Tcc.bib}
 \bibliographystyle{elsarticle-num}

\end{document}